\documentclass[12pt,preprint]{aastex}
\usepackage{multirow}
\usepackage{booktabs}
\usepackage{graphicx}
\usepackage{amsmath}
\usepackage{longtable}
\usepackage{rotating}

\newcommand{\rshadow}{r_{\rm shadow}}
\newcommand{\hrim}{H_{\rm rim,opaque}}

\newcommand{\rsub}{r_{\rm sub}}
\newcommand{\tsub}{T_{\rm sub}}
\newcommand{\hbg}{h_{\rm bg}}
\newcommand{\mdisk}{m_{\rm disk}}
\newcommand{\tmid}{T_{\rm midplane}}

\begin{document}
\title{The Effects of Self-Shadowing by a Puffed up Inner Rim in Scattered Light Images of Protoplanetary Disks}

\shorttitle{Self-Shadowing in Protoplanetary Disks}

\shortauthors{Ruobing Dong}

\author{Ruobing Dong\altaffilmark{1}}
\affil{Lawrence Berkeley National Lab, Berkeley, CA 94720; rdong2013@berkeley.edu}
\affil{Department of Astronomy, University of California, Berkeley}
\altaffiltext{1}{Hubble Fellow}

\clearpage

\begin{abstract}

We explore whether protoplanetary disks with self-shadowing from puffed up inner rims exhibit observable features in scattered light images.  We use both self-consistent hydrostatic equilibrium calculations and parameterized models to produce the vertically puffed up inner rims. We find that, in general, the transition between the shadowed and flared regions occurs in a smooth manner over a broad radius range, and no sudden jump exists at the outer edge of the shadow in either the disk temperature or density structures. As a result, a puffed up rim cannot create sharp ring/arc/spiral-arm-like features in the outer disk as have been detected in recent direct NIR imaging of disks. On the other hand, if the puffed up rim has a sharp edge in the vertical direction, the shadowing effect can produce a distinct 3-stage broken power law in the radial intensity profile of the scattered light, with 2 steep surface brightness radial profiles in the inner and outer disk joined by a shallow transition region around the shadow edge. These types of scattered light profiles may have already been observed, such as in the recent Subaru direct imaging of the TW Hydrae system.

\end{abstract}

\keywords{protoplanetary disks --- stars: pre-main sequence --- stars: variables: T Tauri, Herbig Ae/Be --- circumstellar matter --- radiative transfer --- planets and satellites: formation }


\section{Introduction}\label{sec:intro}

Flattened, rotating circumstellar disks reprocess light from newly born stars in T Tauri and Herbig Ae/Be systems \citep{appenzeller89,waters98}. Disks can modify the spectral energy distribution (SED) of the systems, and reveal themselves in resolved images at various wavelengths \citep{williams11}. In a disk, dust grains only exist outwards the sublimation radius $\rsub$, where dust reach sublimation temperature $\tsub$, leaving a dust-free hole at the center. Protoplanetary disks are usually optically thick to stellar radiation (with exceptions such as transitional disks, \citealt{espaillat14}). As a result, most part of the disk receives direct starlight only at the surface, under a small grazing angle determined by the shape of the surface \citep{chiang97,takami14}. On the other hand, the entire wall of the rim at $\rsub$ is in full sight of the star, heated by intense stellar radiation. As a result, the rim is vertically puffed up under the hydrostatic equilibrium condition as pressure balances gravity \citep{isella05}. As shown by \citet{dullemond01} and \citet{dullemond04-shadowing}, a puffed inner rim blocks starlight from reaching to the disk behind, thereby creating a shadow in the outer disk. Depending on the disk properties, the outer disk may re-emerge from the shadow at a certain distance, or remain shadowed throughout. The puffed up rim and the self-shadowing effect are more prominent in Herbig Ae/Be stars than in T Tauri stars, as dust sublimation occurs closer to the less luminous and cooler T Tauri stars, and the covering angle of the rim is smaller \citep{dullemond01}.

Self-shadowing by a disk rim is a key component in explaining observations of Herbig Ae/Be stars. In particular, the difference between group I and II objects \citep{meeus01} has been interpreted as that group I disks are flared while group II disks are flat and (completely) self-shadowed by a puffed up rim \citep{dullemond04-shadowing,dullemond01}. Also, a puffed up rim has been involved to explain the light curves of UX Orionis stars \citep{dullemond03}, dust emission features in disks \citep{acke04-dustfeature}, and cooling of the surface layer in disks \citep{brown12}. The change of rim height has been proposed to account for the variability of disk infrared excess \citep{espaillat11, muzerolle09, juhasz07} as well as the overall surface brightness of the disk in scattered light imaging \citep{wisniewski08}. Moreover, an evolutionary sequence has been proposed by \citet{acke04-correlation} and \citet{dullemond04-shadowing}: as the shadow cast by the rim extends further as the amount of small grains decreases due to grain growth and settling, eventually an initially flared group I object evolves into a flat self-shadowed group II object at late stages.

Recently, thanks to a few high angular resolution imaging instruments that have been mounted on 8-meter class ground based mirrors in the past few years, near-infrared (NIR) imaging of polarized scattered light from disks have found structures that may be produced by self-shadowing in disks. \citet{marino15} suggested that the two nulls seen on the ring in the $H$ band polarized intensity (PI) image of the HD 142527 system are shadows cast by an misaligned inner disk (see also \citealt{verhoeff11}, \citealt{avenhaus14}, \citealt{fukagawa13}, \citealt{perez14hd142527}, and \citealt{casassus13}). \citet{grady13} found that the two spiral arms in the MWC~758 system may have shadows in its $H$~band PI images (see also \citealt{garufi14}). In addition, various rings and arcs and gaps in disks have also been suggested to be manifestations of self-shadowing (e.g. HD~144432, \citealt{chen12}; HD~169142, \citealt{quanz13-gap}; HD~163296, \citealt{garufi14}; TW Hydrae, \citealt{akiyama15}; HD~169142, \citealt{momose15}). In these cases, the gap region has been interpreted as being shadowed by puffed up inner disk structures, with the outer gap edge presumed to be the edge of the shadow.

On the theory side, the effect of shadowing have been explored by many in the past through radiative transfer modeling, with the shadow cast by either a puffed up inner rim or hydro features produced by planets \citep[e.g.][]{dullemond01,dullemond02,dullemond04-shadowing,vanboekel05, meijer08,tannirkulam08,jang-condell09,jang-condell12,siebenmorgen12}. However, most works focused on the SED or the thermal emission of the disk. The effects of a puffed up rim in high angular resolution scattered light imaging have not been throughly examined. In particular, whether observed rings and arcs in disks can be the outer edge of shadows remain unanswered. 

In this paper, the properties of the shadow cast by a puffed up inner rim in NIR polarized light imaging are studied through radiative transfer simulations using the \citet{whitney13} code. Disk vertical structures are self-consistently calculated assuming hydro static equilibrium (HSEQ). Synthetic images of disks with either a naturally puffed up rim or a parametrized rim are produced, and their properties are measured and compared with observations. The structure of the paper is as following. The models and MCRT simulations are introduced in Section~\ref{sec:models}; the results are presented in Section~\ref{sec:results}, followed by a summary and discussion in Section~\ref{sec:summary}.


\section{Simulation Setup}\label{sec:models}

MCRT simulations are carried out using the \citet[see also \citealt{whitney03a,whitney03b}]{whitney13} code, which has been used to model protoplanetary disks in the past (e.g. \citealt{hashimoto12,zhu12,dong12cavity,dong12pds70,follette13,grady13}). The simulation setup is largely adopted from \citet{dong12cavity}, and is briefly summarized here. The central star has $M_\star=2.5M_\odot$, $R_\star=2R_\odot$, and $T_\star=10^4$K, typical for a Herbig Ae/Be system. The disk is axisymmetric, and is constructed in spherical coordinates $r$ (radial) and $\theta$ (polar, midplane is at $\theta=0$ and poles are at $\theta=\pm90^\circ$). The grid has 500 cells in $r$ direction with logarithmic spacing and 200 cells in $\theta$ direction with power law spacing to over sample the inner disk and the disk midplane. The inner edge of the disk is at dust sublimation radius $\rsub\sim$0.5~AU, where dust temperature reaches 1600~K. The outer edge of the disk is at 200 AU. The surface density $\Sigma$ follows a power law with radius as
\begin{equation}
\Sigma(r)=\Sigma_0\frac{r_0}{r},
\label{eq:sigma}
\end{equation}
where $\Sigma_0$ is the normalization at $r_0$. The $1/r$ power law has been suggested by both detailed disk structure calculation \citep[e.g.][]{dalessio98} and mm disk observations \citep[e.g.][]{andrews09,andrews10}. The total mass of the disk is assumed to be $\mdisk=0.01M_\star=0.025M_\odot$. A small accretion rate of $10^{-9}M_\odot$/yr is assumed to represent a typical accretion state, and also to provide nominal heating to the shadowed region to prevent the development of a known ripple instability \citep{dullemond00, chiang00}. All simulations are run with 1 billion photon packages. Synthetic $H$ band (1.6~$\micron$) images\footnote{The physical quantity recorded in all images is the specific intensity, or intensity $I_\nu$ for short, which has the unit [mJy~arcsec$^{-2}$], or [ergs~s$^{-1}$~cm$^{-2}$~Hz$^{-1}$~arcsec$^{-2}$]. This quantity is sometimes referred to as ``spectral radiance''.} and SED are produced for each model. All models are assumed to be at 140~pc away. Full-resolution MCRT images are convolved by a circular Gaussian kernel with a full width half max (FWHM) of $0.05\arcsec$ to achieve an angular resolution comparable to observations by Subaru, VLT, and Gemini (the FWHM of an airy disk is $1.028\lambda/D\sim0.04\arcsec$ for a primary mirror with a diameter $D=8.2$~m at $\lambda=1.6~\micron$).

In simulations, photons from the central star are absorbed and reemitted or scattered by the dust in the surrounding disk. The temperature in each grid cell is calculated based on the radiative equilibrium algorithm described in \citet{lucy99}. The anisotropic scattering phase function is approximated using the Henyey-Greenstein function. Polarization is calculated assuming a Rayleigh-like phase function for the linear polarization \citep{white79}. The vertical density structure of the disk is solved through iterations assuming self-consistent HSEQ in the vertical direction $z$, as gas pressure balances gravity:
\begin{equation}
\frac{dp}{dz}=-\rho \frac{z}{r} \frac{GM_\star}{r^2},
\label{eq:hseq}
\end{equation}
where $p$ is pressure and $\rho$ is density. Simulations are iterated on the temperature $T$ calculated from the radiative equilibrium solution of the radiation transfer equation at least 10 times to achieve convergence on the vertical structures \citep{whitney13}. In this study the stellar radiation is the main energy source of the disk (heating from accretion has little effect except in the umbra region right behind the rim), and the gravity from the central star is the sole gravitational source. Note that the vertical structure of the disk (as well as the puffed up rim) may be affected by other mechanisms, such as the gravitational perturbation from an orbiting companion (particularly if on an inclined orbit), or turbulences in the disk. The naturally puffed up rims based on the HSEQ condition in this study do not account for these additional mechanisms.

There are two types of dust grains in the models, the ``small'' grains and the ``big'' grains. The gas-to-dust-mass ratio (including both the small and big grains) is assumed to be 100:1 (therefore the total mass of the dust grains $m_{\rm grains}$ is $2.5\times10^{-4}M_\odot$). The small grains represent primordial dust particles. Their properties are assumed to be the standard interstellar medium (ISM) grains as in \citet{kim94}. These grains contain silicate, graphite, and amorphous carbon, and their size distribution is a smooth power law in the range of 0.02-0.25~$\micron$ followed by a sharp cut off beyond 0.25~$\micron$. The small grains are well mixed with gas as the two are dynamically well coupled\footnote{It is possible that $\micron$-sized grains may still have some degree of settling, depending on the strength of the turbulences in disks that stir up dust grains vertically \citep{dullemond04-dustsettling}. Due to a lack of good understanding of the nature and strength of the turbulences in protoplanetary disk, settling of these sub-$\micron$-sized small grains is not considered in this work.}. The large grains represent an advanced stage of grain growth and coagulations. Model 1 in \citet{wood02} is adopted for their properties. They are made of amorphous carbon and astronomical silicates, and have a power-law size distribution as $n(s)\sim s^{-3}$, an exponential cutoff beyond $50\micron$, and a maxim size of 1 mm. The big grains tend to settle to the disk midplane, as they usually dynamically decouple from the gas and are no longer held in the vertical direction by the gas pressure \citep{dullemond04-dustsettling}. The vertical distribution of the big grains is assumed to be a Gaussian profile with a small scale height $\hbg=0.01r (r/1{\rm AU})^{0.25}$ (i.e. their vertical structure is parametrized and NOT evolved in the HSEQ iterations). Therefor the big grains form a thin layer at the disk midplane. The details of the properties and spatial distribution of the big grains have little effect on the scattered light images and SED presented below, as long as they are well embedded at the disk midplane. The optical properties of both grains can be found in Figure~2 in \citet{dong12cavity}. 

Two sets of models, series A and B, are constructed and listed in Table~\ref{tab:models}. Model Base is the fiducial model, which has no big grains. Series A is constructed by fixing $m_{\rm grains}$ while decreasing $m_{\rm grains,small}$ and increasing $m_{\rm grains,big}$, similar to the ML series in \citet{meijer08} and BL series in \citet{dullemond04-shadowing}. This scenario (from model Base to A4) is to mimic an evolutionary sequence: the amount of small grains decreases with time as grain growth and coagulation gradually convert them into bigger particles that settles down at the disk mid-plane. The inner rim in Base to A4 is naturally puffed up under the HSEQ condition (only the small grains). As will be shown in the next section, while series A does gradually turn from a flared disk (Base) into a self-shadowed disk (A4), the boundary between the shadowed inner disk and the flared outer disk is not sharp. This is because the rim does not have a sharp edge in the vertical density distribution, and the radial optical depth of the rim only gradually decreases with height. To maximize the sharpness of the rim edge and produce features with the highest contrast possible in scattered light imaging, series B is created. In B1 to B6, an artificial opaque rim is placed at $r=\rsub$ to $1.1\ \rsub$, with a parametrized height $\hrim$ expressed in the angle subtended by half of the rim to the star (i.e. the vertical structure of the artificial rim is fixed and NOT evolved under the HSEQ condition). The opaque rim is constructed by filling the region with small grains of a uniform density $10^{-9}\ {\rm g\ cm}^{-3}$, so that the radial optical depth $\tau$ of the rim is greater than $10^4$ at wavelengths of interests, therefore completely blocks the starlight in the solid angle subtended by the rim. In other words, in the polar direction $d\rho/d\theta=0$ inside the rim, then $\rho=0$ outside the rim. This artificial vertically constant density structure is only designed in order to create the sharpest possible shadows. It is not meant to have a physical motivation. Both the naturally puffed up rim and the parametrized artificial rim are illustrated in Figure~\ref{fig:verticalprofiles}.


\section{Results}\label{sec:results}

This section presents the main modeling results, first the structure of the disks (Section~\ref{sec:structure}) and then observational aspects of the models (Section~\ref{sec:modelobservations}).

\subsection{Disk Structure}\label{sec:structure}

\begin{figure}
\begin{center}
\includegraphics[trim=0 0 0 0, clip,width=0.48\textwidth]{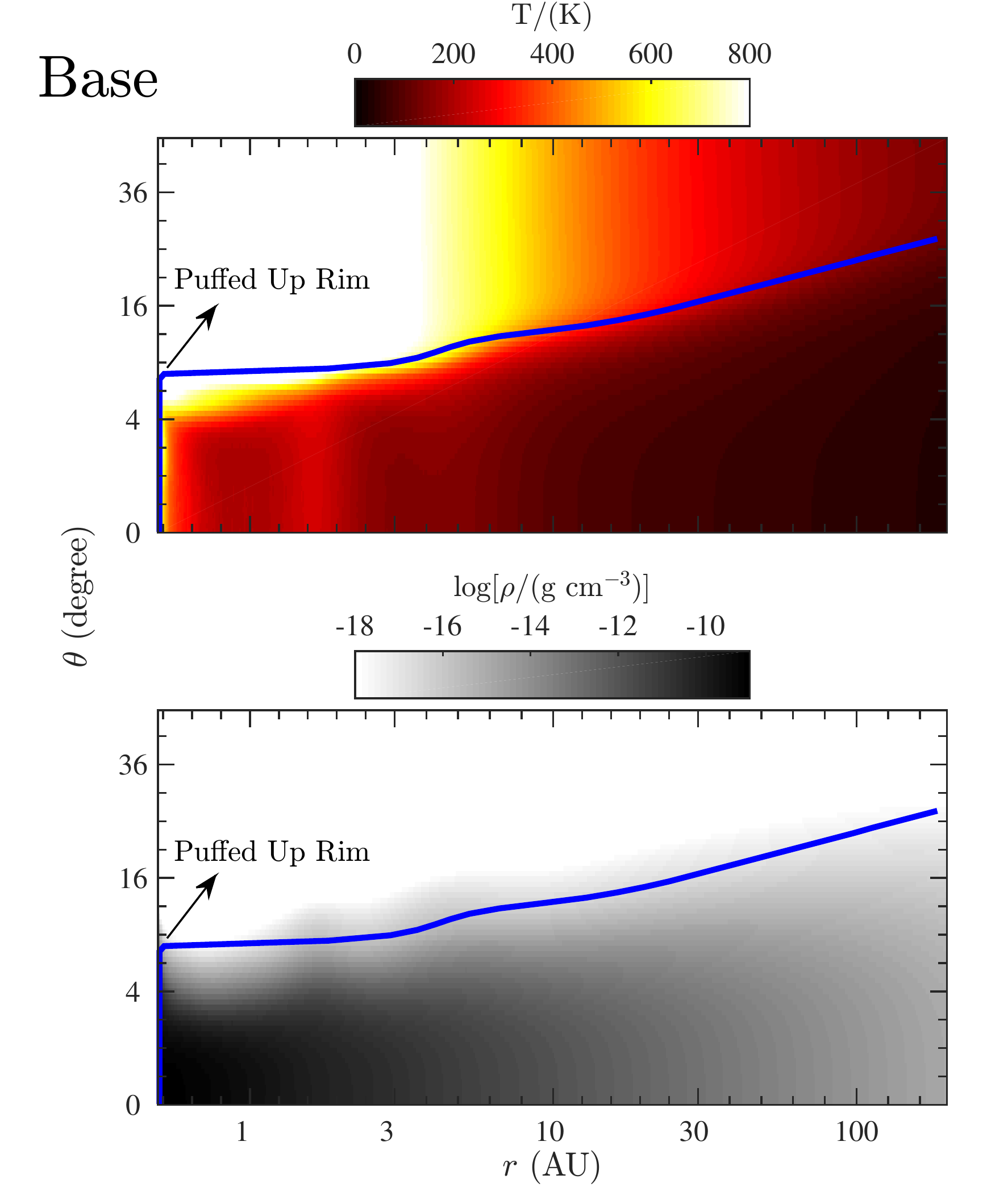}
\includegraphics[trim=0 0 0 0, clip,width=0.48\textwidth]{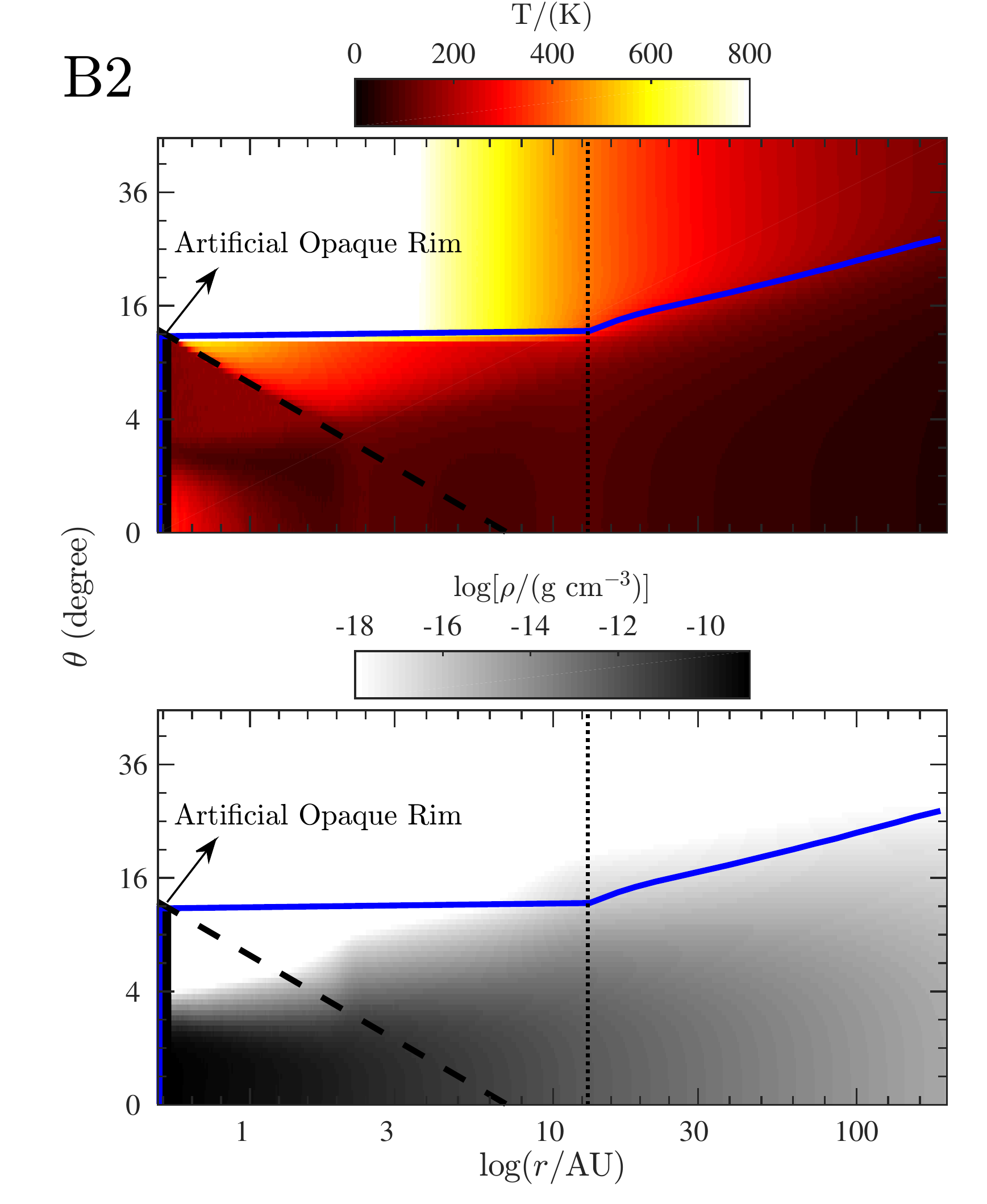}
\end{center}
\figcaption{2D density ($\rho$) and temperature ($T$) structure as a function of radius $r$ and polar angle $\theta$ in spherical coordinate for model Base (left column) and B2 (right column). The blue line in each panel marks the disk surface, defined as where the radial optical depth $\tau$ at 0.4~$\micron$ at each $\theta$ angle from the star reaches 1 (the stellar flux peaks around 0.4~$\micron$). The puffed up rim in model Base and the parametrized artificial rim in model B2 at $\rsub$ are labeled. The region under the dashed line in model B2 is the umbra, where the entire star is not visible to the disk (above this line the polar region of the star may be visible). The vertical dotted line in B2 marks the edge of the shadow at $\rshadow=13$~AU. The $\tau=1$ surface traces out the boundary between the hot disk atmosphere and the cool interior; the umbra region is sufficiently cold so that the disk collapses to the midplane; while the $\tau=1$ surface and the umbra boundary further separate out a region that receives a small amount of direct stellar radiation from the polar region of the star and is moderately puffed up in the vertical direction.
\label{fig:verticalprofiles}}
\end{figure}

The density and temperature structure for models Base and B2 are shown in Figure~\ref{fig:verticalprofiles} in spherical coordinates, with the disk surface over-plotted, defined as where the radial optical depth $\tau$ at 0.4~$\micron$ at each polar angle $\theta$ from the stellar surface (the stellar flux peaks aroundt 0.4~$\micron$) reaches 1. In spherical coordinates a shadowed region has a flat surface (constant $\theta$ with increasing $r$), while a flared region has a rising surface (increasing $\theta$ with increasing $r$). The disk surface traces out the boundary between the hot disk atmosphere and the cool interior, as most starlight intercepted by the disk is deposited near the surface. 

In the model Base, a puffed up inner rim is clearly visible in the density structure as a vertically extended structure close to $\rsub$. The rim casts a shadow behind from $\rsub$ to $\sim3$~AU, where the surface comes out of the shadow and starts rising. However, as discussed in Section~\ref{sec:models}, the puffed up rim has a blurry edge. As a result, the edge of the shadow is not sharp, and the disk surface smoothly transits from flat to flared around 3~AU.

In the model B2, a parametrized artificial rim is placed at $\rsub$ with $\hrim=12.5^\circ$. The sharp rim edge results in a sharp turning point on the $\tau=1$ surface at $\rshadow=13$~AU, where the surface suddenly rises. Inside the shadow region, the disk does not have a uniform structure, and can be separated into two regions. An umbra exists immediately behind the rim (boundary indicated by the dashed line), within which the entire surface of the star is not visible. In the umbra, practically the only heating sources are radiation diffusion and accretion. This region becomes sufficiently cold to collapse to the midplane. On the other hand, the region in between the umbra and the $\tau=1$ surface receives additional heating from the following ways:
\begin{enumerate}
\item Once outside the umbra and moving towards $\rshadow$, the disk receives direct radiation from a bigger and bigger polar region on the star.
\item Inward heat diffusion from $\rshadow$ heats up the adjacent region, smoothing out the transition and preventing a sudden jump in $\tmid$ \citep{dullemond01}.
\item Indirect illumination and heating from the halo \citep{dullemond04-shadowing}. In particular, closer to $\rshadow$, secondary (multiple) scattering from the disk atmosphere in the flared region  becomes more important.
\item NIR flux from the opposite side of the inner rim.
\end{enumerate}
The strength of additional heating sources 1-3 increases with radius, puffs up the disk moderately starting at $\sim$2~AU. Eventually the $\tau=1$ surface comes out of the shadow at 13 AU and disk stays flared outward.
 
\begin{figure}
\begin{center}
\includegraphics[trim=0 0 0 0, clip,width=0.49\textwidth]{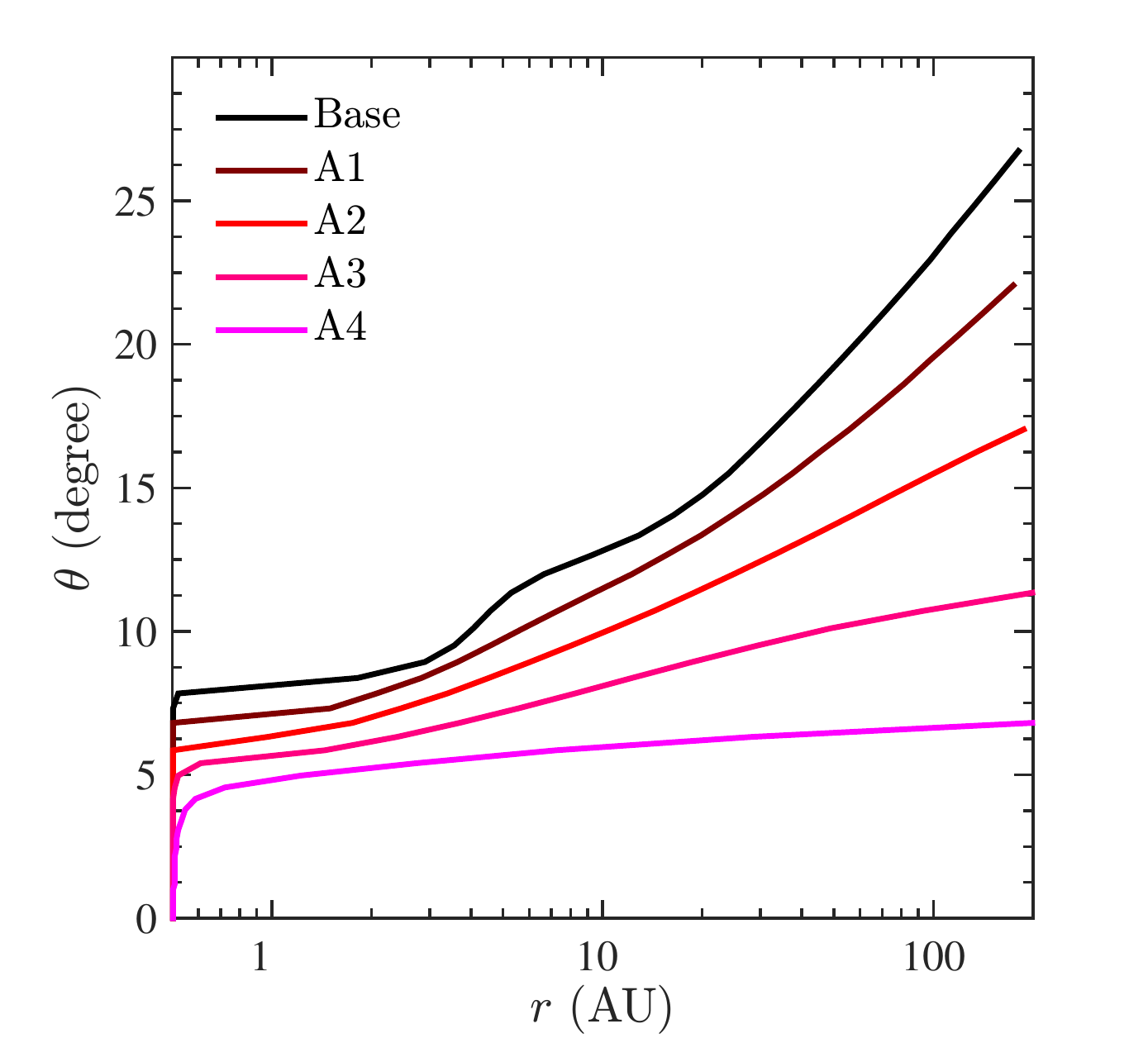}
\includegraphics[trim=0 0 0 0, clip,width=0.49\textwidth]{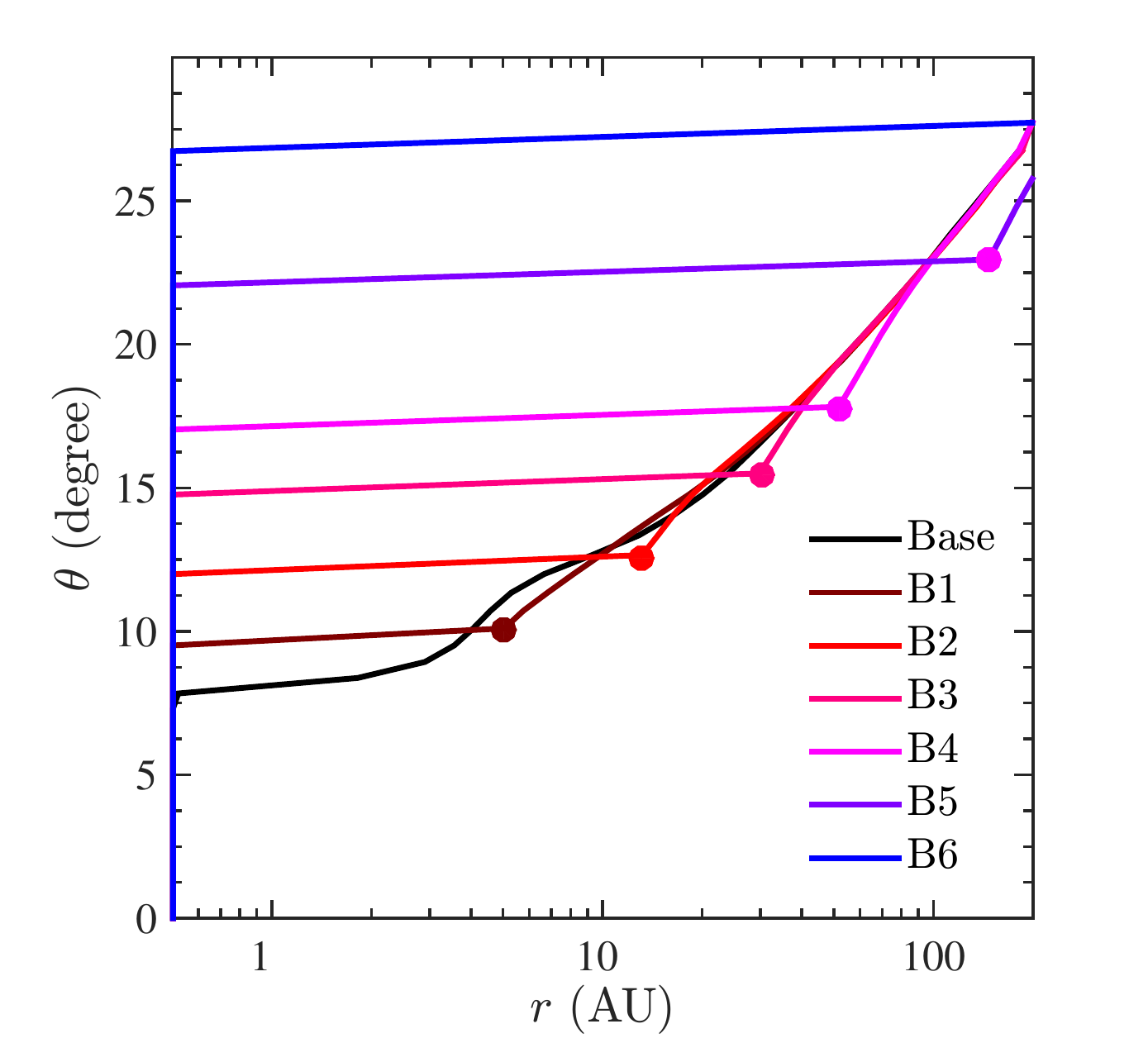}
\end{center}
\figcaption{Radial $\tau=1$ surface in spherical coordinates. A shadowed disk has a flat surface (constant $\theta$ with increasing $r$) while a flared disk has a rising surface (increasing $\theta$ with increasing $r$). In the A series the shadow edge is difficult to define, as the disk transits from the self-shadowed region to the flared region in a smooth way, due to a lack of sharp rim edge. In the B series the transition point is obvious, marked by the dot at [$\rshadow,\hrim$] on each curve, where the surface comes out the shadow cast by the inner rim and starts to rise. Model B6 stays in the shadow of its high rim throughout the disk. This surface transition point rises and moves outward as the height of the artificial rim increases, and the disk surface always converges to model Base beyond $\rshadow$, indicating that the flared part of the disk in all models are similar.
\label{fig:surface}}
\end{figure}

The $\tau=1$ surface for all models are plotted in Figure~\ref{fig:surface}. As shown in \citet{dullemond04-shadowing} and \citet{meijer08}, from Base to A4 the disk gradually evolves from flared to flat (completely self-shadowed by its own puffed up rim), as the amount of small grains decreases. The shadowed-to-flared-disk transition occurs more and more smoothly, and $\rshadow$ cannot be clearly defined. On the other hand, the surface transition is sharp in the B series, as the surface suddenly turns upward at [$\rshadow$,$\hrim$] (marked by dots in the right panel and labeled in Table~\ref{tab:models}. The entire disk in B6 is inside the shadow so $\rshadow$ is not defined.). The transition point rises and moves outward as $\hrim$ increases, and the surface always converges to model Base once out of the shadow, indicating that the flared part of the disk in all models are similar.

\begin{figure}
\begin{center}
\includegraphics[trim=0 0 0 0, clip,width=0.48\textwidth]{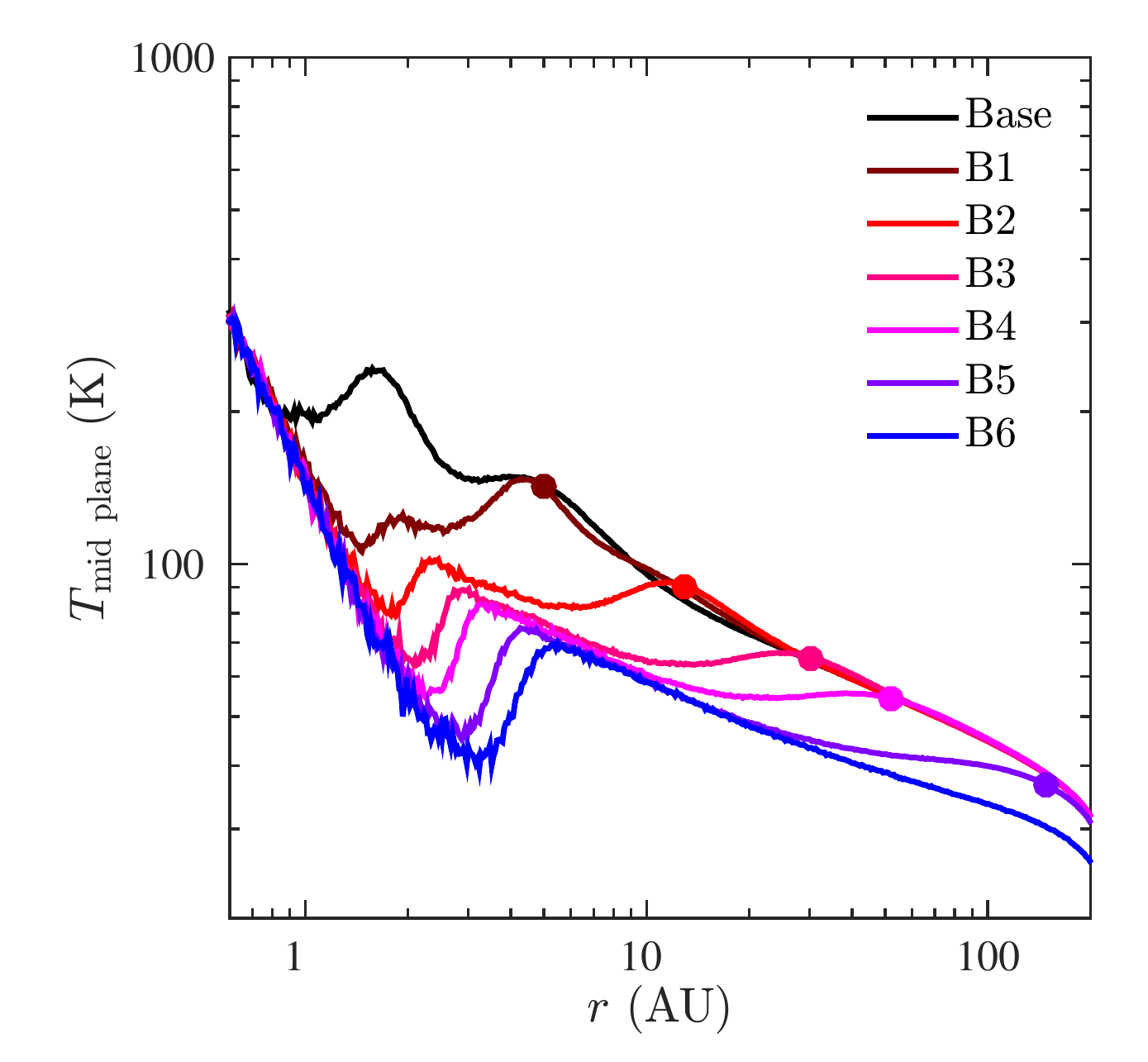}
\end{center}
\figcaption{Mid-plane temperature profile for the B series. The shadow edge at $\rshadow$ is marked by the dots on the curves. There are two bumps on the curves. The first bump occurs when the disk comes out of the umbra, and the second bump occurs around the shadow edge $\rshadow$. The transition of $\tmid$ around $\rshadow$ is smooth, without a sudden jump.
\label{fig:midplanet}}
\end{figure}

The midplane temperature of the disk $\tmid$ is shown in Figure~\ref{fig:midplanet} for the B series. There are two bumps on the curves. The first bump occurs when the disk comes out of the umbra, and the second bump occurs at the shadow edge $\rshadow$. In the umbra all models have the same $\tmid$ profile (best traced out by model B6). The location of the first bump moves outward as $\hrim$ increases and the umbra extends. After the first bump the shadow reduces the midplane temperature, consistent with \citet{siebenmorgen12}, and models converge again to another profile in cases that the two bumps are not too close to each other (B3 to B6). Finally, all models converge to the Base model at $r>\rshadow$, echoing the convergence of the $\tau=1$ surface in Figure~\ref{fig:surface}, and demonstrating the similarity of the models in the flared region. The $\tau=1$ surface officially comes out of the shadow at $\rshadow$ and starts to be directly irradiated by the bulk part of the stellar surface ($|$latitude$|>\hrim$). However, the transition to the ``flared profile'' occurs much earlier than $\rshadow$. For example, B4 starts to deviate from its ``shadow profile'' (traced out by B6) at $\sim$10 AU, and gradually converges towards its ``flared profile'' (traced out by Base) until the eventual merger at $\rshadow=52$~AU. This echoes the smooth transition in temperature and density structures seen in Figure~\ref{fig:verticalprofiles}.

Vertical density structure of the disk is determined by the temperature structure under the HSEQ condition. Therefore, at radius of a fraction of $\rshadow$, the disk already starts to deviate from its thin ``shadow state'' and become modestly puffed up, until eventually smoothly joins the flared disk at $\rshadow$. This is illustrated in Figure~\ref{fig:verticalprofiles} for B2, in which case the thin ``shadow state'' ends at $\sim$2~AU, and in between 2~AU and $\rshadow=13$~AU the disk is modestly puffed up, and beyond 13~AU the disk becomes fully flared.

In summary, the disk in Series A evolves from a mostly flared disk to a self-shadowed flat disk, but the shadowed-to-flared-disk transition is smooth, due to the lack of a sharp rim edge. In the extreme cases, with parametrized sharp rims in series B, the transition on the $\tau=1$ surface becomes sharp, but the transition in the underlying disk density and temperature structure is still smooth and occurs over a radius range.

\subsection{Scattered Light Images and SED}\label{sec:modelobservations}

\begin{figure}
\begin{center}
\includegraphics[trim=0 0 0 0, clip,width=0.48\textwidth]{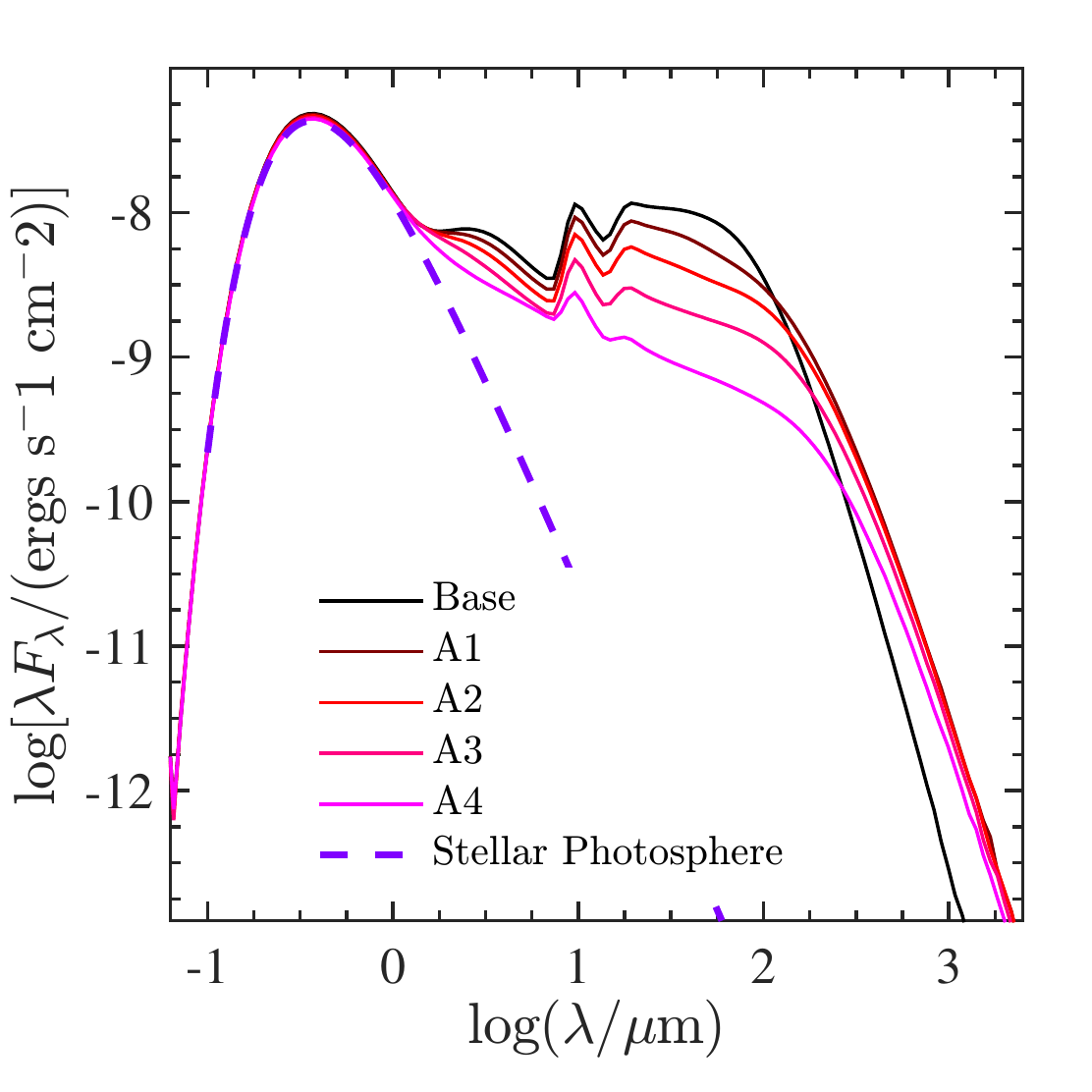}
\includegraphics[trim=0 0 0 0, clip,width=0.48\textwidth]{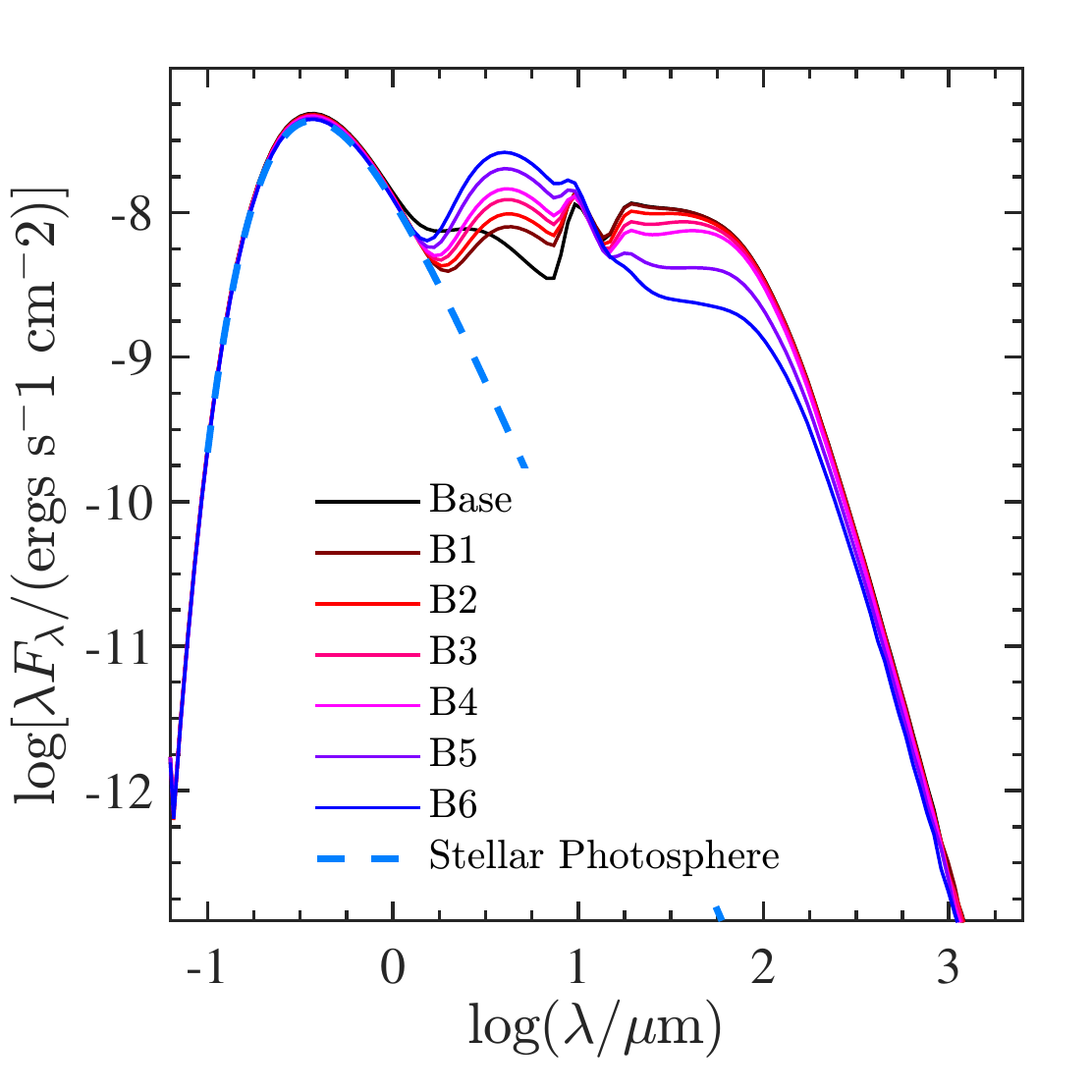}
\end{center}
\figcaption{SED of the models at $45^\circ$ inclination.
\label{fig:sed}}
\end{figure}

SED for all models are shown in Figure~\ref{fig:sed}. Series A mimics the evolutionary sequence from group I to II, as shown in \citet{dullemond04-shadowing}. The 1-100 $\micron$ infrared excess drops as the total amount of small grains decreases and the disk becomes more and more self-shadowed, meanwhile the far infrared and mm emissions beyond 100~$\micron$ rise due to more big grains. For series B, the NIR excess from 1-10~$\micron$ comes from the artificial rim, which increases as the rim grows higher from B1 to B6. Meanwhile the 10-100~$\micron$ excess decreases as the shadow edge expands and a bigger disk region falls into the shadow. This is similar to the ``seesaw'' phenomenon in \citet{espaillat11}, which has been explained by puffed up rims with varying heights. Beyond 100~$\micron$ the emissions mainly come from the outer part of the disk, which remains flared and has a similar density and temperature structure in most models thus the emission is little changed.

Scattered light images trace structures at the disk surface, set by the dust distribution and intrinsically determined by the temperature structure under the HSEQ condition. While formally inside $\rshadow$, the shadowed region may still receives and scatters NIR photons from (1) the polar region of the star (in the region immediately below the radial $\tau=1$ surface); (2) the bulk part of the star (high in the disk atmosphere); (3) the opposite side of the rim; and (4) secondary scattering photons from the disk halo (in particular at radius close to $\rshadow$). If the rim is naturally puffed up, stellar photons may also be scattered by the tenuous rim edge into the shadowed region. Therefore, it is expected that the shadowed region to be not completely dark, and the surface brightness should converge to the flared disk when getting closer to $\rshadow$.

\begin{figure}
\begin{center}
\includegraphics[trim=0 0 0 0, clip,width=0.5357\textwidth]{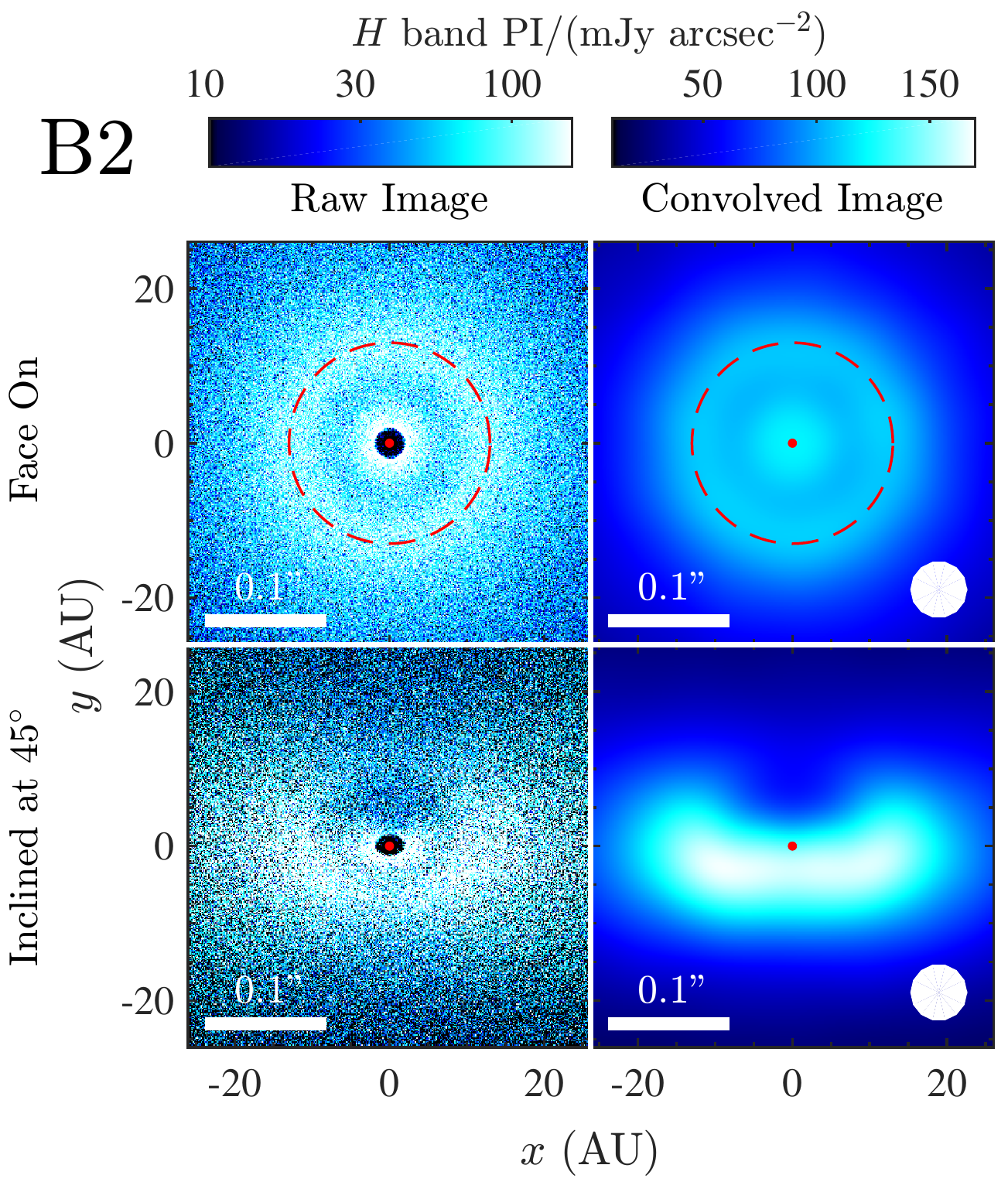}
\includegraphics[trim=0 0 0 0, clip,width=0.45\textwidth]{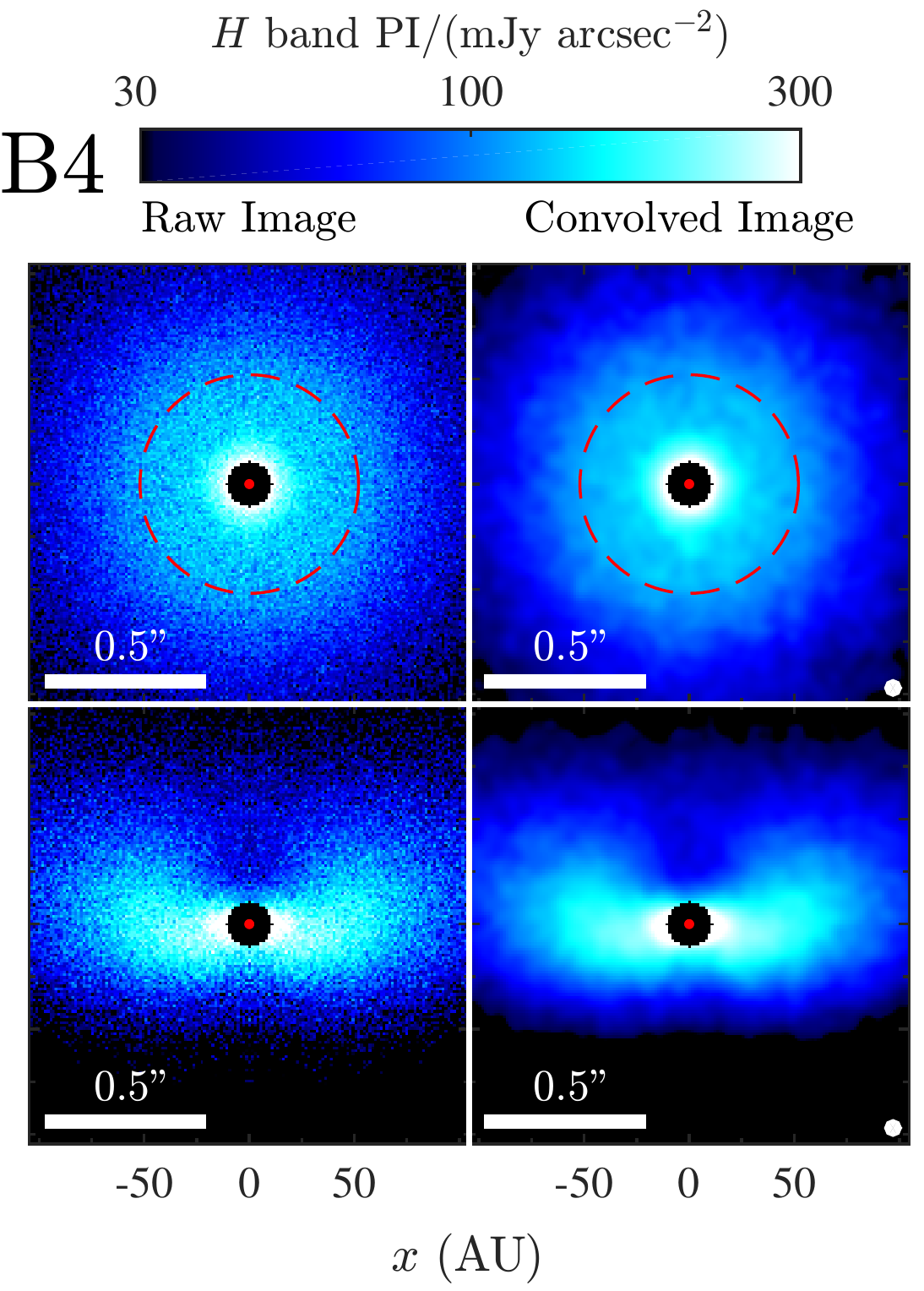}
\end{center}
\figcaption{$H$~band polarized intensity images for B2 and B4, showing full resolution raw images (left column in each model) and images convolved by a Gaussian PSF with FWHM=0.05$\arcsec$ (right column in each model) at face-on (top row) and 45$^\circ$ inclination viewing angles (bottom row, lower half is the front side). Models are assumed to be at 140~pc away. The dashed circle in face-on images marks $\rshadow$. The red dot in all panels marks the location of the star. The scattered light from the artificial inner rim is removed. The central bright region in B4 is masked out for better illustration of the outer disk. The PSF size is indicated at the lower right corner in the convolved images. B4 and raw images of B2 are shown in logarithmic scale, while convolved images in B2 are in linear scale to highlight the shadow edge feature. The ring/gap structure at $\rshadow$ in B2 is marginally traceable, while it is not visible in B4.
\label{fig:images}}
\end{figure}

\begin{figure}
\begin{center}
\includegraphics[trim=0 0 0 0, clip,width=0.49\textwidth]{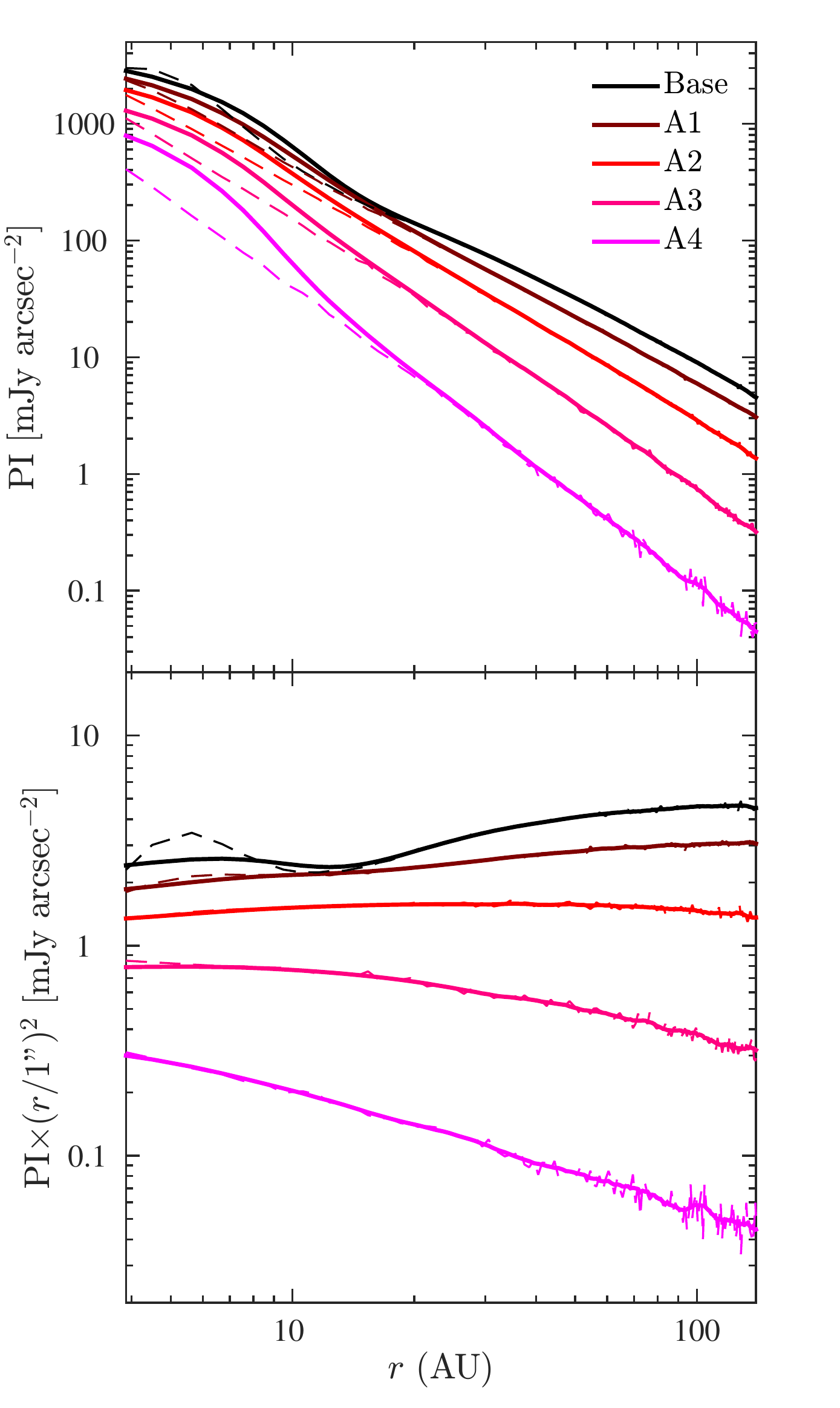}
\includegraphics[trim=0 0 0 0, clip,width=0.49\textwidth]{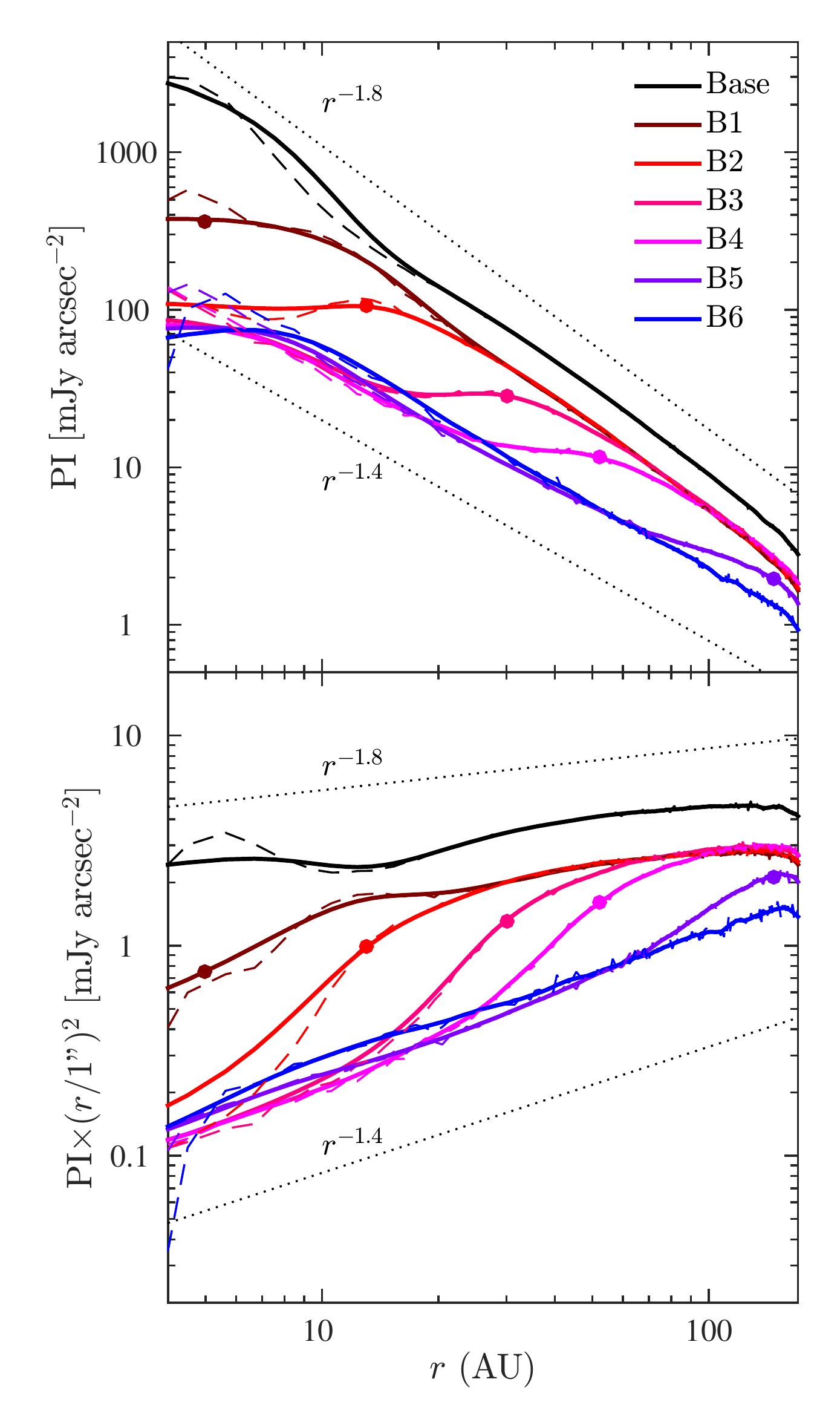}
\end{center}
\figcaption{Surface brightness as a function of radius for face-on $H$~band PI images for all models. Upper panels show the intrinsic surface brightness, and lower panels show the $(1/r)^2$ scaled values. Solid curves are convolved images (Gaussian PSF FWHM$=0.05\arcsec$; targets at 140~pc); dashed curves are full resolution images; and dotted lines indicate the trend lines. $\rshadow$ is marked for B1 to B5 as a dot. The A models always have a smooth radial profile reasonably well characterized by a single power law, while the general surface brightness decreases by 1-2 orders of magnitudes from Base to A4.. On the other hand the B series are characterized by a 3-stage broken power law, with two steep radial profiles in the inner and other disk and a shallow one in between around $\rshadow$. Even in the extreme cases with the sharpest possible rim edge in the B models, the shadow edges do not manifest themselves as prominent rings in images. Merely they create a region in between the inner shadowed disk and the outer flared disk with a shallow radial profile close to $r^0$. 
\label{fig:image-rp}}
\end{figure}

$H$~band polarized intensity images for model B2 and B4 are shown in Figure~\ref{fig:images}. In B2, a marginal ring/gap structure (i.e. a region with PI$\sim r^{>0}$) at $\rshadow$ can be seen at face-on viewing angle, inside which the shadowed region has low surface brightness. If we define ``ring/gap contrast'' as the ratio of the azimuthally averaged surface brightness at the peak of the ring to the bottom of the gap, the contrast is 1.33 for the raw image and 1.03 for the convolved image. At a 45$^\circ$ inclination angle, the ring becomes elliptical, and due to the forward/backward scattering effect the front side is brighter than the back side \citep{hashimoto12}. In B4, the shadow edge is not distinctive in all panels. At face-on viewing angle, the azimuthally averaged surface brightness decreases with radius throughout the disk (see Figure~\ref{fig:image-rp}).

The radial profile of face on $H$~band PI images for all models are shown in Figure~\ref{fig:image-rp}. For the A models, converting small grains into vertically settled big grains gradually turns a bright flared disk into a faint flat one. The brightness of the disk drops by 1-2 orders of magnitude. However this process happens in a rather smooth manner that the surface brightness of the entire disk drops all together, while the outer disk drops faster than the inner region. The dependence of PI on $r$ changes from PI$\sim r^{-1.8}$ in the Base model to PI$\sim r^{-2.5}$ in A4. The surface brightness distribution remains smooth and relatively featureless throughout in all cases. In particular, no ring/gap structure exists at the edge of the shadowed region, in either the original images or the $1/r^2$ scaled images.

For the B models, the shadowed-to-flared-disk transition is clearly visible on the radial profiles. In general, the surface brightness radial profile can be described by a 3-stage broken power law: (1) well inside the shadow different models have a similar radial profile as PI$\sim r^{-1.4}$ (best seen in B3 to B6 as their shadows extend far enough); (2) in the flared region all models converge to PI$\sim r^{-1.8}$, the same as the Base model; and (3) a transition region with PI~$\sim r^0$ (constant PI) connects the two. This is consistent with the trend seen earlier in the disk density, temperature, and surface structures, that B models have essentially the same structure both in the inner shadowed region and the outer flared region, with a wide transition region in between. And similar to the $\tmid$ profile, the transition occurs over a radius range and not suddenly. For B2 to B5\footnote{$\rshadow$ in B1 is too close to the star and beyond the capability of current NIR imaging facilities.}, radial profiles smoothly deviate from the $r^{-1.4}$ profile around $\rshadow/2$, and gradually converge to the $\sim r^{-1.8}$ profile around $\rshadow$. B3 has a marginal ``gap/ring'' structure similar to B2, with very weak ring/gap contrast ratios (listed in Table~\ref{tab:models}). PI never increases with radius throughout the disk in B4 to B6, and a ring/gap structure is not present around the shadow edge. The difference between B2/B3 and B4/B5 is probably caused by the fact that the transition region is wider in B4/B5 due to their large shadow size, so the transition occurs even more smoothly.


\section{Summary and Discussion}\label{sec:summary}

The effects of puffed up inner rims in protoplanetary disks in both temperature and density structure and NIR scattered light imaging are studied. Two types of rims are explored: natural rims that are self-consistently puffed up under the hydrostatic equilibrium condition (series A), and artificial opaque rims with parametrized heights and sharp edges in the vertical direction (series B). The main results from the models are:
\begin{enumerate}
\item An inner rim naturally puffed up under the HSEQ condition creates a shadow behind, within which the disk surface is flat and beyond which it is flared (the A models). Decreasing the amount of small grains gradually turn the entire disk into a self-shadowed and flat disk, confirming previous results such as in \citet{dullemond01,dullemond04-shadowing,meijer08} (Figure~\ref{fig:surface}). 
\item For naturally puffed up rims, the shadowed-to-flared-disk transition occurs smoothly, and no sharp/distinct turning point on the disk surface exists. This is mainly due to the fact that HSEQ rims do not have a sharp vertical edge, and their ``radial optical depth'' gently decreases with height. As a result, the scattered light images have a smooth/featureless surface brightness distribution. From Base to A4, the surface brightness of the disk drops by 1-2 orders of magnitude, and the radial dependence changes from a shallower $r^{-1.8}$ to a steeper $r^{-2.5}$ (Figure~\ref{fig:image-rp}).
\item The most extreme rims, which are parametrized artificial rims with a razor sharp vertical edge in the B series, do create a sharp turning point on the radial $\tau=1$ surface (Figure~\ref{fig:surface}). This point formally defines the edge of the shadow $\rshadow$, beyond which the surface is flared and sees the bulk part of the star. Both the inner shadowed region and the flared region are similar in models with different rim heights. On the other hand, the shadowed-to-flared-disk transition does not occur abruptly in the underlying disk density and temperature structures. The midplane temperature starts to deviate from its ``shadow profile'' at a fraction of $\rshadow$ (Figure~\ref{fig:midplanet}), and eventually converges to its ``flared profile'' around $\rshadow$. The density structure of the disk follows the change in $\tmid$ under the HSEQ condition, thereby exhibits a broad transition region as well (Figure~\ref{fig:verticalprofiles}). This is mainly caused by partial illumination by the polar region of the star, multiple scattering, and radiation diffusion within the disk.
\item Even the parametrized razor sharp rims in the B models cannot produce prominent rings and arcs in scattered light images as what have been directly imaged recently. On the other hand, self-shadowing by the rim does have a clear effect on the radial profile of the scattered light. The surface brightness radial profiles of the B models can be described by a 3-stage broken power law: two steep profiles in the inner shadowed region and the flared region, connected by a shallow profile in a broad transition region (Figure~\ref{fig:image-rp}). The enhancement around the shadow edge however is usually not strong enough to make a prominent ring/gap structure. B2 and B3 show a very marginal ring/gap structure around $\rshadow$ (Figure~\ref{fig:images}), while other models remain largely smooth and featureless throughout in both the original and $1/r^2$ scaled images.
\end{enumerate}

Remarks on the implications of the models and observations:
\begin{enumerate}
\item A faint disk in NIR direct imaging observations with a relatively steep surface brightness radial profile (i.e. $r^{\lesssim-2}$) is consistent with small grains being heavily depleted and the entire disk is self-shadowed by the inner rim. The surface brightness threshold depends on the stellar luminosity. For a $10^4$~K, 2.5 $M_\odot$, and 2 $R_\odot$ central star, the threshold is about 1/$(r/{\rm 140\ AU})^2$ mJy arcsec$^{-2}$. Such kind of faint disks may have already been found in observations (e.g. HD~139614, Subaru direct imaging, M. Fukagawa, private communication, Hashimoto et al. in prep.).
\item Even with the sharpest possible rims as the parametrized rims in B models, the shadow edge fails to manifest itself as prominent rings/arcs in scattered light images. The highest ring/gap contrast ratio in full resolution images is only 1.33, and practically no ring/gap structure is found in convolved images with an angular resolution of 0.05$\arcsec$. Note that the convolved images in this study have the ideal chance to review potential disk structures, as (1) they are convolved by a Gaussian PSF, which smears the intrinsic full resolution images less than a realistic PSF that has a similar kernel but more extended wings; (2) no observational noises and instrument effects are included, which are expected to further weaken the detectability of features in images. 
\item Realistic rims in A models can only produce structures even smoother than B models in scattered light images. Radial structures with high contrast recently observed in NIR direct imaging surveys, such as the arcs in the HD~163296 system \citep{garufi14}, the gap in the HD~169142 system \citep{momose15}, the spiral arms in the SAO~206462 system \citep{muto12,garufi13} and the MWC~758 system \citep{grady13,benisty15}, and various rings and arcs in transitional disks \citep[e.g.][]{hashimoto12,mayama12} are unlikely to be the shadow edges of puffed up rims. It is more likely that they are created by local density or scale height structures \citep{juhasz15}, such as gaps \citep[e.g.][]{zhu11,dodsonrobinson11,dejuanovelar13,pinilla15,dong15a} and spiral arms (e.g. Dong et al. 2015b) produced by planets in the density distribution\footnote{An inclined inner disk can produce azimuthally confined shadows with high contrast on the outer disk, as shown in \citet{marino15}.}.
\item Though unable to create arcs/rings and gaps in disks, a shadow cast by a puffed up rim may still have an effect on the surface brightness radial distribution of scattered light images. With a vertically sharp rim, scattered light images show a 3-stage broken power law: 2 steep radial profiles in the inner and other disk connected by a shallow transition region around the shadow edge. Such systems may have already been observed: the $H$~band scattered light image of TW Hydrae has a shallow $r^{-0.3}$ transition region located at $r\sim$20-40~AU, in between 2 steep $r^{-1.4}$ and $r^{-2.7}$ profiles in the inner and outer disk \citep{akiyama15}. The shallow region may actually be a gap, which can be either confirmed or ruled out by future observations with better image quality and S/N ratio. If the shadow region is not a gap, then comparing with the B models this may suggest that the disk is self-shadowed, and the shadow edge is around $\sim30$~AU (note that TW Hydrae has an $r\sim4$~AU inner hole at the center, \citealt{calvet02,hughes07}, therefore perhaps the shadow is created by a puffed up cavity wall at $\sim4$~AU instead of a rim at $\sim0.5$~AU). However a self-consistent structure puffed by the HSEQ condition is unlikely to account for the observation, as the shadowed-to-flared-disk transition is too smooth. The needed puffed up structure has to have a somewhat sharp cutoff, perhaps sculpted by other mechanisms in the system. The realization of such a structure with a physical origin can be the subject of future studies.
\end{enumerate}


\section*{Acknowledgments}

I thank Misato Fukagawa, Jun Hashimoto, and Barbara Whitney for useful discussions and help on the paper, and the referee Cornelis Dullemond for a helpful referee report. This work is supported by NASA through Hubble Fellowship grant HST-HF-51320.01-A awarded by the Space Telescope Science Institute, which is operated by the Association of Universities for Research in Astronomy, Inc., for NASA, under contract NAS 5-26555. This research made use of the SAVIO cluster at UC Berkeley, and the Lawrencium cluster at the Lawrence Berkeley National Laboratory (Supported by the Director, Office of Science, Office of Basic Energy Sciences, of the U.S. Department of Energy under Contract No. DE-AC02-05CH11231).



\clearpage


\begin{table}
\footnotesize
\begin{center}
\caption{Model Properties}
\begin{tabular}{cccccccc}
\tableline\tableline
Model  & Parametrized/Natural Rim \tablenotemark{a} &  $f_{\rm grains,small}$\tablenotemark{b}  & $\hrim$ ($^\circ$)\tablenotemark{c} & $\rshadow$ (AU)\tablenotemark{d} & Ring/Gap Contrast\tablenotemark{e} \\
\tableline
Base & Natural & 1 & N/A & N/A & N/A \\
\tableline
A1 & Natural & $10^{-1}$ & N/A & N/A & N/A \\
A2 & Natural & $10^{-2}$ & N/A & N/A & N/A\\
A3 & Natural & $10^{-3}$ & N/A & N/A & N/A\\
A4 & Natural & $10^{-4}$ & N/A & N/A & N/A\\
\tableline
B1 & Parametrized & 1 & 10 & 5 & N/A  \\
B2 & Parametrized & 1 & 12.5 & 13 & 1.33/1.03 \\
B3 & Parametrized & 1 & 15 & 30 & 1.10/1.02 \\
B4 & Parametrized & 1 & 17.5 & 52 & N/A \\
B5 & Parametrized & 1 & 22.5 & 147 & N/A \\
B6 & Parametrized & 1 & 27.5 & N/A & N/A \\
\tableline
\end{tabular}
\tablecomments{Properties of the models. (a) Whether the inner rim is a naturally puffed up rim (i.e. the vertical structure of the rim is self-consistently determined by the HSEQ condition) or a parametrized artificial opaque rim (i.e. the rim structure is fixed and NOT evolved under the HSEQ condition; see Section~\ref{sec:models} for details). (b) The mass fraction of the sub-$\micron$-sized small grains in the model $m_{\rm grains,small}/m_{\rm grains,total}$ (the rest of the grains are up to mm-sized big grains). (c) the height of the artificial rim expressed in the angle extended by half the rim to the star. (d) The edge of the shadow cast by the rim, defined as the turning point on the $\tau=1$ surface (Figure~\ref{fig:surface}). A distinct shadow edge does not exist in the A models and B6. (e) The contrast of the ring/gap feature around $\rshadow$ in face-on $H$~band PI image, defined as the ratio of the azimuthally averaged surface brightness at the peak of the ring to the bottom of the gap inside. Only model B2 and B3 has such a feature (i.e. a region with PI$\sim r^{>0}$). The 2 numbers are for full resolution raw images (left) and convolved images (right, Gaussian PSF FWHM=$0.05\arcsec$).  $\rshadow$ in B1 is too small to have a meaningful ring/gap definition for current NIR imaging facilities.}
\label{tab:models}
\end{center}
\end{table}

\end{document}